# How Spammers and Scammers Leverage AI-Generated Images on Facebook for Audience Growth

*Much of the research and discourse on risks from artificial intelligence (AI) image generators, such as DALL-E and Midjourney, has centered around whether they could be used to inject false information into political discourse. We show that spammers and scammers—seemingly motivated by profit or clout, not ideology—are already using AI-generated images to gain significant traction on Facebook. At times, the Facebook Feed is recommending unlabeled AI-generated images to users who neither follow the Pages posting the images nor realize that the images are AI-generated, highlighting the need for improved transparency and provenance standards as AI models proliferate.*

Renee DiResta (1) & Josh A. Goldstein (2)
Affiliations: (1) Stanford Internet Observatory, Stanford University, United States, (2) Center for Security and Emerging Technology, Georgetown University, United States
Draft: March 2024

## Research questions
- How are profit and clout-motivated Page owners using AI-generated images on Facebook?
- When users see AI-generated images on Facebook, are they aware of the synthetic origins?

## Essay summary
- We studied 120 Facebook Pages that posted at least 50 AI-generated images each, classifying the Pages into spam, scam, and 'other creator' categories. Some were coordinated clusters of Pages run by the same administrators. As of March 5, 2024, the Pages had a mean follower count of 128,877 and a median follower count of 71,000.
- These images collectively received hundreds of millions of engagements and exposures. A post including an AI-generated image was one of the 20 most viewed pieces of content on Facebook in Q3 2023 (with 40 million views and more than 1.9 million interactions).
- Spam Pages used clickbait tactics and attempted to direct users to off-platform content farms and low-quality domains. Scam Pages attempted to sell products that do not exist or to get users to divulge personal details; some were posting the AI-generated images on stolen Pages.
- The Facebook Feed (formerly "News Feed") at times shows users AI-generated images even when they do not follow the Pages posting those images. We suspect that AI-generated images appear on users' Feeds because the Facebook Feed ranking algorithm promotes content that is likely to generate engagement. Facebook has increased the percentage of "unconnected posts" (posts from Pages that users do not follow) that appear in users' Feeds over the last three years. Media coverage has reported that engaging with AI-generated images often results in users receiving recommendations for more AI-generated image content; this was our anecdotal experience with our own Feeds as well.
- Comments on the AI-generated images suggest that many users are unaware of their synthetic origin, though a subset of users post comments or infographics alerting others and warning them of scams. Viewer misperceptions highlight the importance of labeling and additional transparency measures moving forward.
- Some of the Pages in our sample that posted unlabeled AI-generated images also used known deceptive practices, such as account theft or takeover, and exhibited suspicious follower growth.

## Implications

With the diffusion of new generative AI tools, policymakers, researchers, and the public have expressed concerns about impacts on different facets of society. Existing work has developed taxonomies of misuses and harm (Weidinger et al., 2022; Ferrara 2024) and tested the potential of AI tools for generating instructions for biological weapons (Mouton et al., 2024), propaganda (Spitale et al., 2023; Goldstein et al., 2024), and phishing content (Grbic & Dujlovic 2023). A significant portion of this literature is theoretical or lab-based, and focused on political speech, such as impacts on elections and threats to democracy and shared capacity for sensemaking (Seger et al., 2020).

And yet, even in the realm of the political, the tactics of manipulators have long been previewed by those with a different motivation: making money. Spammers and scammers are often early adopters of new technologies because they stand to profit during the time gap between when technology makes novel, attention-capturing tactics possible and when defenders recognize the dynamics and come up with new policies or technological interventions to minimize their impact (e.g. Metaxas & DeStefano, 2005; Goldstein & DiResta 2022). Recall the Macedonian teenagers behind the high-profile "fake news" debacles of 2016: investigative reports found that they used eye-catching content—promoted by Facebook's recommendation algorithm—to drive users to off-platform websites where they would collect advertising revenue via Google Adsense (e.g. Subramanian 2017; Hughes & Waismel-Manor 2021).

While the misuse of text-to-image and image-to-image models in politics is worthy of study, so are deceptive, non-political applications of new technologies. Understanding misuse can shape risk analysis and mitigations. In this article, we show that images from AI models are already being used by spammers, scammers, and other creators running Facebook Pages, and, at times, achieving viral engagement. For the purposes of this study, we describe Pages as "spam" if they post low-effort (e.g. AI-generated or stolen) content at high frequencies AND (b) use clickbait tactics to drive people to an outside domain or (c) have inauthentic follower growth (e.g. from purchasing fake accounts). We categorized Pages as "scam" if they (a) deceive followers by stealing, buying, or exchanging Page control, (b) falsely claim a name, address, or other identifying feature, and/or (c) sell fake products. Our 'other creator' category includes Pages that post AI-generated images at high frequency and are not transparent about the synthetic origin of content, but we do not have clear evidence of manipulative behavior. Developing a fuller taxonomy for scam, spam, and other deceptive Pages falls outside our remit: our goal here is simply to make explicit the types of Pages described in the study.

We studied 120 spam, scam, and other creator Facebook Pages that shared large numbers of AI-generated images to capture the attention of Facebook users. Many of these Pages formed clusters, such as 6 Pages with more than 400,000 collective followers that declared themselves affiliated with the "Pil&Pet Corporation" in the Intro section of their Pages. The Pil&Pet Pages posted AI-generated images with similar captions and had Page operators from Armenia, the U.S., and Georgia. Some Pages we studied did not overtly declare a mutual affiliation, but they posted on highly similar topics, recycled posts, and co-moderated Facebook Groups or shared links to the same domains. Others were not clearly connected to others in the list, but used highly similar captions, identical generated images, or images on similar themes. A number of Pages, for example, posted AI-generated images of log cabins. At times, these AI-generated images were recommended to users via Facebook's Feed (including in our own Feeds). The posts are not transparent about the use of AI and many users do not seem to realize that they are of synthetic origin (Koebler 2023).

*Consequences and Recommendations*

The Pages we studied may produce both direct and indirect problematic consequences. In terms of direct consequences, we observed unambiguously manipulative behaviors from some of the page operators, such as page/account theft and leveraging batches of inauthentic followers to enhance their legitimacy or to engage in comment discourse with content viewers. AI-generated content appears to be a boon for spam and scam actors because the images are easy to generate, often visually sensational, and attract engagement. In terms of indirect consequences, most of the AI-generated images we saw from these Pages did not include an indication of their synthetic origin. We observed that Facebook users would often comment on the pictures in ways suggesting they did not recognize the images were fake—congratulating, for example, an AI-generated child for an AI-generated painting. Scam accounts occasionally engaged with credulous commenters on the posts, both in Pages and Groups, at times seeking personal information about them or offering to sell them products that do not exist. Journalists have likewise found ghost kitchens advertising AI-generated food on DoorDash and Grubhub (Maiberg 2024) and AI-generated images of fake goods on shopping websites (Ma et al., 2023). The increasing complexity of distinguishing between real and synthetic content online will likely further exacerbate issues with trust in media and information..

To combat the deceptive use of AI-generated images, there are steps different actors can take. First, social media companies should invest resources in detecting both scams as well as AI-generated content. For the latter, they should test the effect of different interventions for labeling AI-generated content (including labeling images they detect, requiring users to label images upon uploading, and rolling out watermarking techniques). Researchers should investigate whether tech companies are true to the voluntary commitments announced at the 2024 Munich Security Conference for grappling with AI-generated content (e.g. "attaching provenance signals to identify the origin of content where appropriate and technically feasible") (Munich Security Conference, 2024). Second, the media, and AI generation tool creators themselves, should help the public understand AI image generation tools in a manner that is digestible and not sensational. This could include Public Service Announcement clips that help people understand that AI-generated images can look photorealistic. Such announcements should learn from recent work on inoculation theory (e.g. Roozenbeek 2022) and teach proactive user strategies (e.g. lateral reading). Third, researchers should contribute to understanding the effects of AI-generated content on broader information landscapes. For example, studies can build on earlier work to examine whether labeling content as AI-generated will increase trust in information that is not labeled (Jakesch et al., 2019). Other work could take an ethnographic approach and interview individuals behind these Facebook Pages to better understand their motivations and views about transparency and deception.

## Findings

*Finding 1: Spammers, scammers, and other creators are posting unlabeled AI-generated images that are gaining high volumes of engagement on Facebook. Many users do not seem to recognize that the images are synthetic.*

Unlabeled AI-generated images from the Pages we studied amassed a significant number of views and engagements on Facebook. One way that we discovered pages deceptively using AI-generated images was by observing repeated caption text across accounts—even accounts that were seemingly unconnected. For example, AI-generated content of old people, amputees, and infants often contained the phrase "No one ever blessed me" in the caption. AI-generated images of people alongside their supposed woodworking, ice sculptures, or drawings contained variants on the phrase "Made it with my own hands" in the caption; neither the person depicted nor the piece of art are real. Phrases such as "This is my first cake! will be glad for your marks" explicitly solicit feedback from users (see Figure 1).

Oftentimes these posts would receive comments of praise or congratulations in response. Sometimes the post text made little sense in the context of the presented image; an AI-generated image of Jesus rendered as a crab worshiped by other crabs also proudly declared "Made it with my own hands!" and received 209,000 engagements and more than 4,000 comments (see Figure 2).

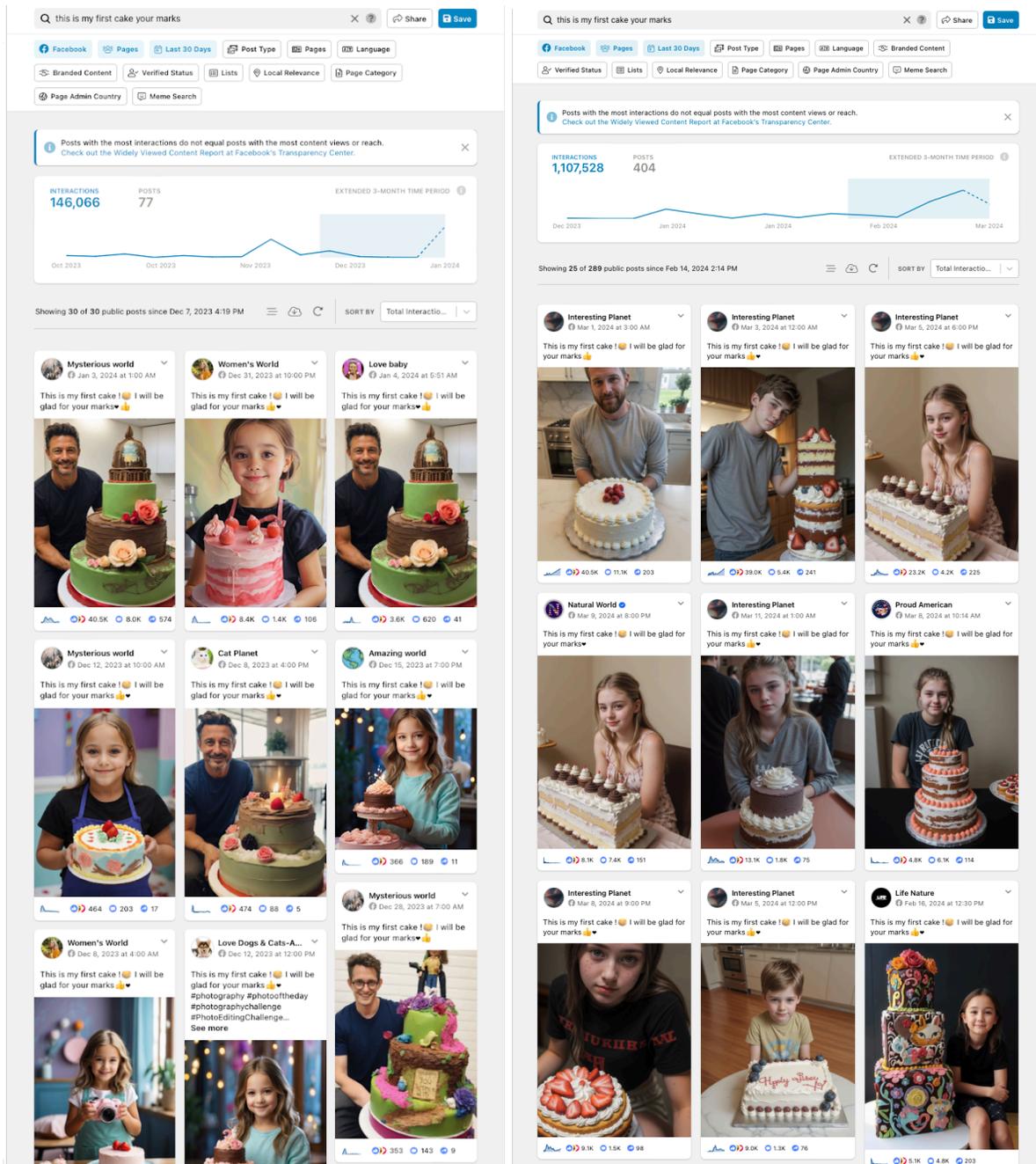

***Figure 1. Screenshots from CrowdTangle (a public insights tool from Meta) showing interactions with posts with the phrase 'This is my first cake your marks.'*** *The screenshot on the left from January 2024 shows a variety of Pages posting AI-generated cake photos. The screenshot on the right, taken in March 2024, shows that posts with the phrase garnered more than a million collective interactions since December 2023.*

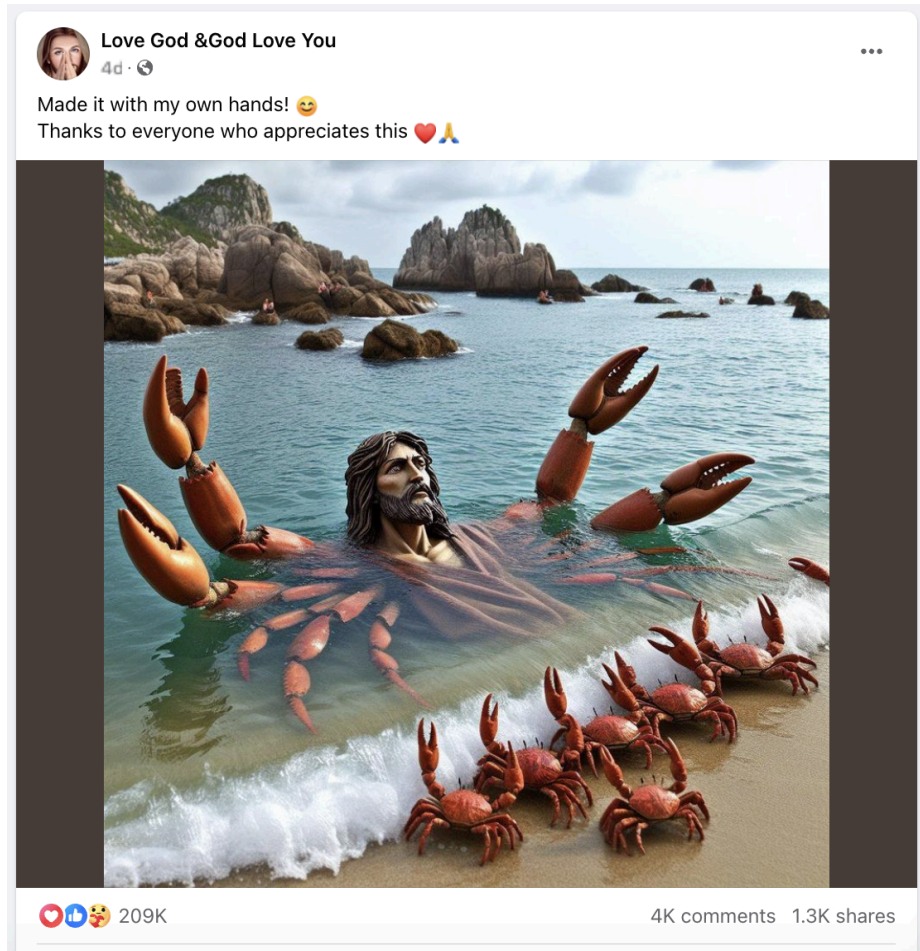

*Figure 2. Screenshot of a Page sharing an AI-generated image of crab Jesus.* The example shows how incredulous some of the claims are; some commenters mention that directly.

Common themes for content across the Pages we studied included at least 43 Pages posting AI-generated houses or cabins, 25 posting AI-generated images of children, 17 posting AI-generated wood carvings, and 10 posting AI-generated images of Jesus. We provide examples of other AI-generated images posted by Pages in the dataset with high levels of engagement in Figure 3.

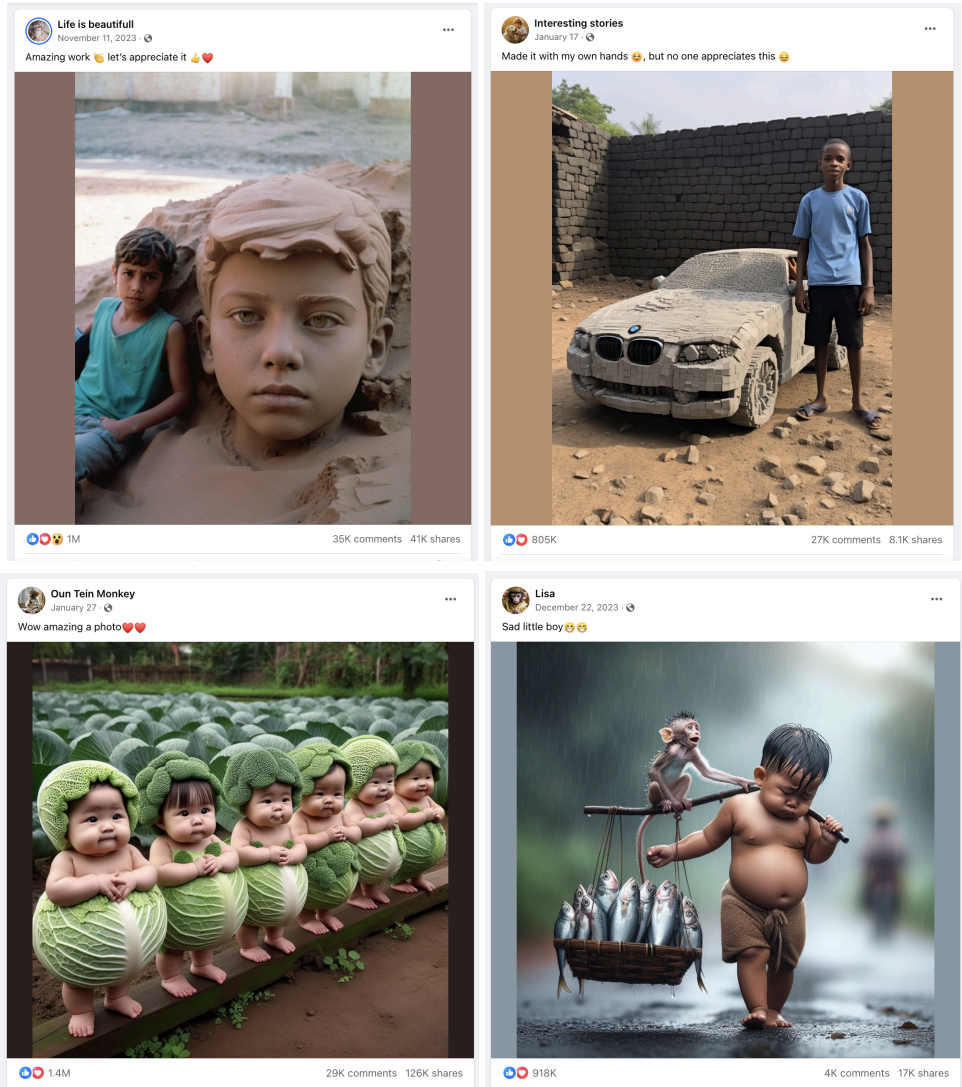

*Figure 3. Screenshots of AI-generated images posted by Pages in the dataset that receive large numbers of engagements (top left: 1 million; top right: 805,000; bottom left: 1.4 million; bottom right: 918,000). The posts do not indicate that the images are AI-generated.*

While researchers typically can view the number of engagements a post has (the sum of reactions, comments, and reshares), they do not have access to the number of views for a given post. This is not the case, however, for the 20 most viewed pieces of content in a given quarter, which are listed on Facebook Transparency's Center. One of the 10 most viewed posts in Q2 2023 was an unlabeled AI-generated image from a Page that transitioned from a cooking Page to one showing AI-generated images of kitchens (Figure 4).

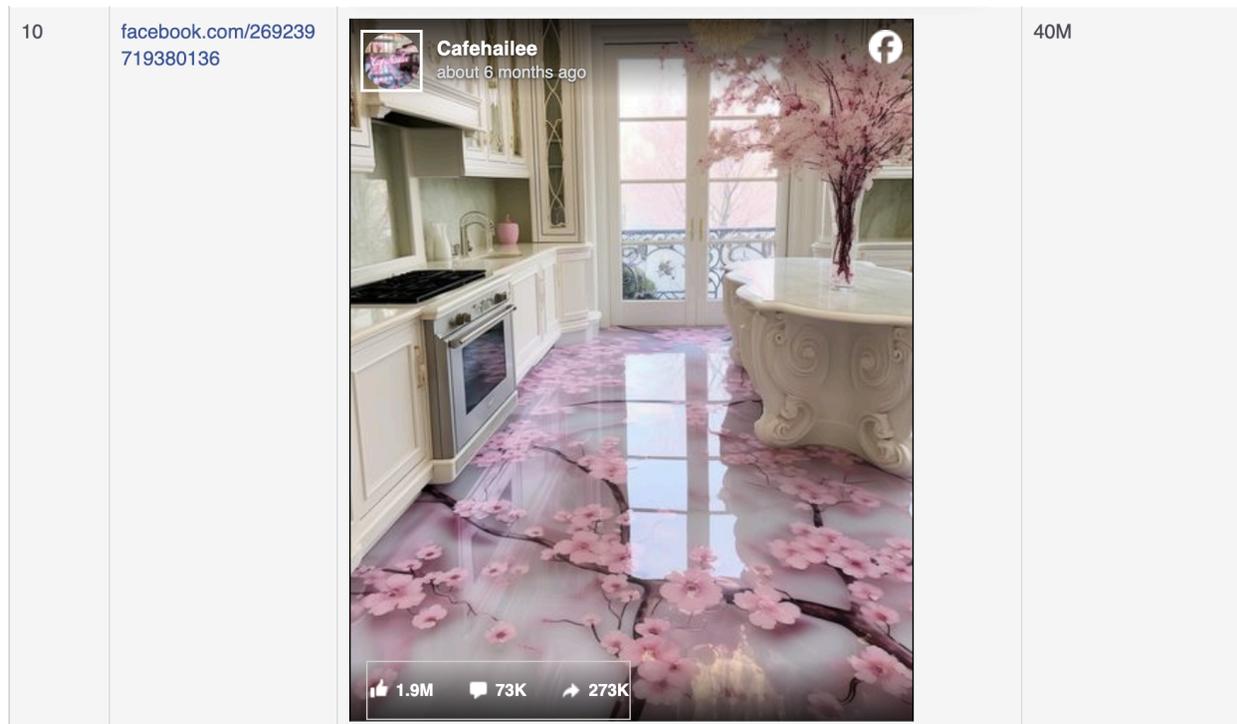

*Figure 4. Screenshot from Facebook's Transparency Center Q3 2023 report, showing a post from the Facebook Page "Cafehailee."* The post received 40 million views and 1.9 million engagements. It has signs of being AI-generated, including small abnormalities in the stove knobs and cabinet handles.

*Finding 2: The Facebook Feed at times recommends unlabeled AI-generated images to users who do not explicitly follow the Page posting the content.*

We suspect these high levels of engagement are partially driven by the Facebook recommendation algorithm. In 2022, Alex Heath reported on an internal memo by Facebook President Tom Allison about planned changes to the algorithm that would "help people find and enjoy interesting content regardless of whether it was produced by someone you're connected to or not." According to Heath, it was clear to Meta that to compete with TikTok, it had to compete with the experience of TikTok's main "For You" Page, which shows people content based on their past viewing habits and anticipated preferences (independent of whether the user follows the creator's account) (2022).

Each quarter, beginning in 2021, Meta publishes the "Widely Viewed Content Report: What People See on Facebook." The report includes a section that breaks out where posts in Feed come from (e.g. from Groups people had joined; content their friends had shared; from sources they are not connected to, but Facebook thinks they might be interested in, etc.). We pulled the portion of Feed views from different sources, reported each quarter from Q2 2021 (when Facebook began publishing it) through article drafting in Q3 2023. As shown in Figure 5, the portion of content views from "unconnected posts" (posts from Pages that users do not follow) from Facebook algorithm recommendations rose dramatically from 8% in Q2 2021 to 24% in Q3 2023.

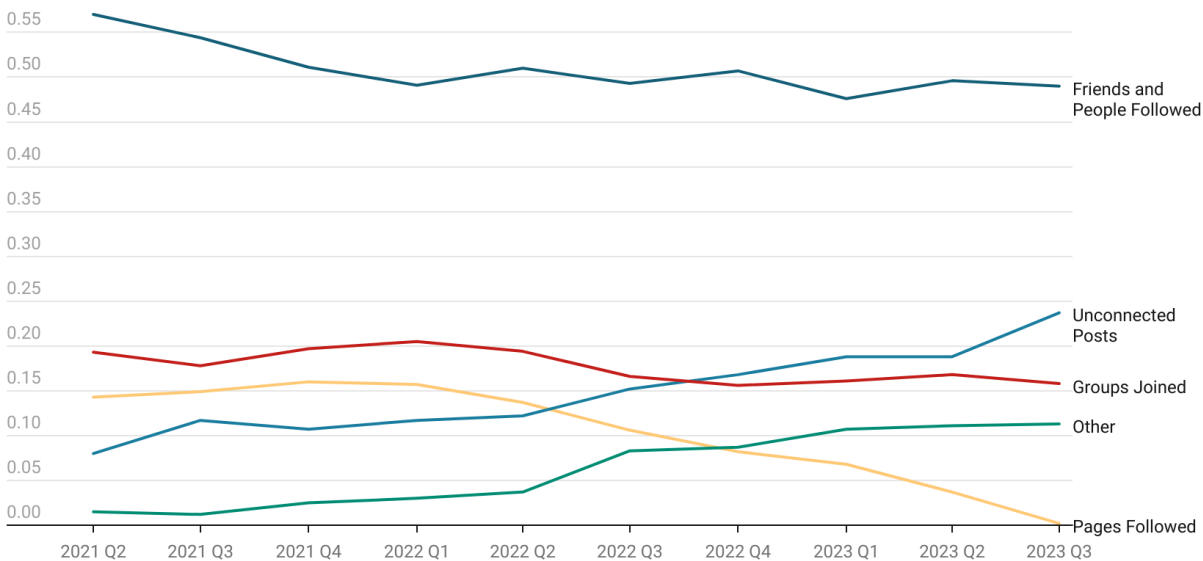

*Figure 5. Portion of Facebook Feed views from different sources.* Unconnected posts rose from 8% of Feed views in 2021 Q2 to 24% in 2023 Q3. Figure created using Datawrapper.

After we conducted preliminary research, we began to see more and more of AI-generated images in our social media Feeds, despite not following or liking any of the Pages posting AI-generated images. The algorithm likely expected us to view or engage with AI-generated images because we had clicked on others in the past. Two colleagues who reviewed our work reported that they were shown AI-generated images in their Feed before they even began investigating, and we observed a number of social media users claiming large influxes of AI-generated images in their Facebook Feeds (Koebler 2024b). For example, Reddit users are discussing their Facebook Feeds with comments such as "Facebook has turned into an endless scroll of Midjourney AI photos and virtually no one appears to have noticed."

*Finding 3: Scam and spam Pages leveraging large numbers of AI-generated images are using well-known deceptive practices, such as Page theft or repurposing, and exhibit suspicious follower growth.*

Researchers of social media influence operations have often discovered the use of tactics, techniques, and procedures to obtain accounts with large followings. Obtaining a Page with an existing following provides a ready-made audience that can be monetized, while obtaining a fake audience can boost follower counts thereby increasing the perceived credibility of a Page (Phua and Ahn 2016). Some of the Facebook Pages we followed used tried-and-true tactics along these lines: 46 had changed their names, often from an entirely different subject, and some displayed a massive jump in followers after the name change (but prior to new activity that would organically have produced that follower spike).

Take, for example, the Page "Life Nature." The Page was first created on December 9, 2011, with the name "Rock the Nation USA." The Pages appears to be the Page of a real band, posting fliers for the traveling band with information about upcoming concerts. On December 29, 2023, the Page changed its name to "Life Nature" and began posting AI-generated images (among photos taken from other parts of the Internet). Whereas the touring band had ~9,400 followers, a number which had remained consistent from July 2023 to December of that year, following the name change the Page acquired 300,000 followers (December 31 to January 6). The second post after the name change received more than 32,000 likes and 17,000 comments. Figure 6 shows changes in content from band posters to AI-generated content. A booking agent for the band told 404 Media's Jason Koebler, who found the Page as part of his own investigation, "we found out about the page being hacked towards the end of December. No idea how it happened unfortunately, as I was the only admin and my personal profile is still intact. Appreciate you trying to support the cause" (Koebler 2024a). Figure 7 shows the increase in follower growth. Since our analysis, the Page is no longer live on Facebook.

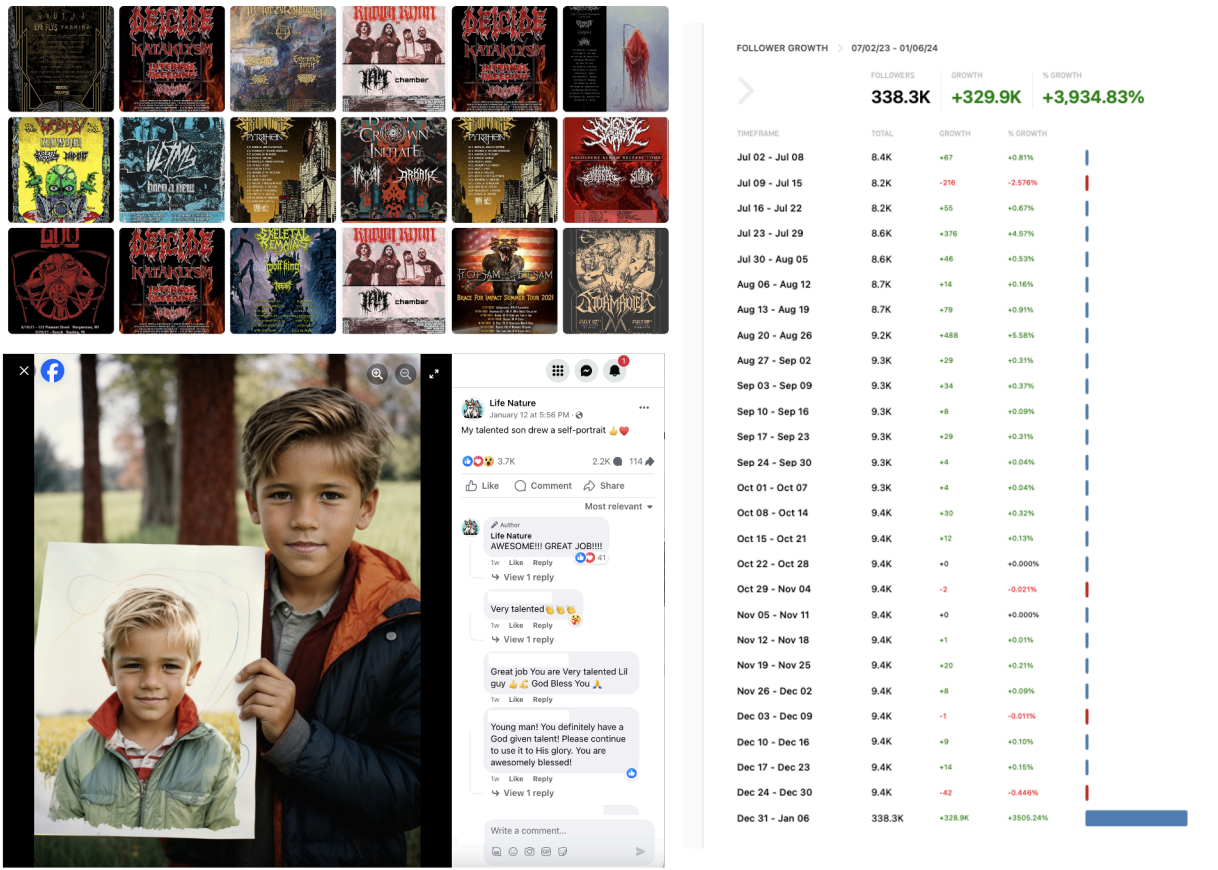

*Figure 6. Photos posted by the Page "Life Nature" prior to its name change (top left) and an AI-generated image posted by the Page after the name change (bottom left). Screenshots from CrowdTangle showing follower growth for the Facebook Page "Life Nature" (right).* The Page received 328,900 new followers from December 31, 2023 to January 6, 2024.

Other examples of Facebook Pages that posted a large number of AI-generated images but were either stolen or repurposed include "Olivia Lily" (formerly a church in Georgia), "Interesting stories" (formerly a windmill seller), and "Amazing Nature" (formerly equine services).

Spam Pages largely leveraged the attention they obtained from viewers to drive them to off-Facebook domains, likely in an effort to garner ad revenue. They would post the AI-generated image often using overlapping captions as described in Finding 1, then leave the URL of the domain they wished users to visit in the first comment under the image. For example, a cluster of Pages that posted images of cabins or tiny homes pointed users to a website that purportedly offered instructions on how to build them. Other clusters used AI-generated or enhanced images of celebrities, babies, animals, and other topics to grab attention, and then directed users to heavily ad-laden "content farm" domains—some of which themselves appeared to be primarily articles composed of AI-generated text. A look at posting dynamics of several Pages created prior to the advent of easily-available generative AI tools suggested that they both increased their posting volume and also transitioned from posting primarily links to their domains with clickbait titles, to posting attention-grabbing AI-generated images instead (see Appendix Figure 1). This is potentially due to the perception that the Facebook recommendation engine was likely to privilege one content type over another.

Scam Pages used images of animals, homes, and captivating designs as well, but then often implied that they sold the product. Users that appeared to be fake profiles (new accounts; stolen and reversed profile photos featuring minor online celebrities) engaged with commenters about the potential to purchase the product or obtain more information about it. These behaviors were distinct from other high-posting-volume AI-generated image Pages that also appeared to be capitalizing on AI-generated content for audience growth, including some that ran political ads, but which were not demonstrably manipulative.

*Finding 4: A subset of Facebook users realized that the images were AI-generated, and took steps to warn other users.*

While most comments on the AI-generated images were unrelated to the artificial nature of the images, some users recognized Pages relying on AI-generated images and engaging in other suspicious behavior.

For example, take the Page "Love Baby." From November 2019 through June 2021, the majority of Reviews on the Page described their positive experiences visiting a store in Maryland. They talked about the holidays and supporting local businesses. However, recent reviews include no mention of the store but rather "mostly Fake/AI" (November 20, 2023) or "all contents are AI GENERATED, so fake" (January 17, 2024). The change in reviews correspond to a likely change in Page control, as the Page—which included profile pictures of Catonsville Mercantile—until May 22, 2023, then included profile pictures of baby photos or other unrelated pieces of content.

Others posted comments on photos from the Page claiming the content is AI-generated. Similar to scam alert comments on other Pages, these comments often include infographics that claim the Pages are engaged in nefarious activities like identifying targets for scams or scraping information about Facebook users.

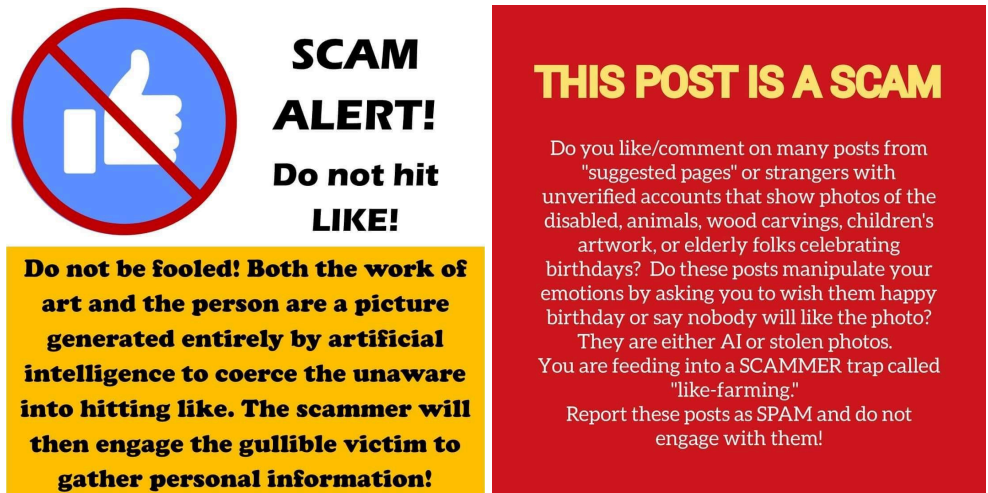

*Figure 8. Example of "Scam Alert" infographics posted as comments on posts from the "Love Baby" Page that likely include AI-generated photos.*

During the time of our investigation, Meta announced its plan to roll out labeling of AI-generated images that it could detect (Clegg 2024). We did not find many photos labeled yet. However, as shown in Figure 9, we did find at least one image labeled as "False information." The label linked to an article from Congo Check highlighting that the image was AI-generated and engaging in comment-bait, or encouraging interaction to artificially increase engagement and reach (Watukalusu 2024). The article cited the high number of engagements with the post; it was likely fact-checked because it went viral. Similar images from other Pages are not labeled, showing the difficulty of scaling fact-checking of AI-generated images.

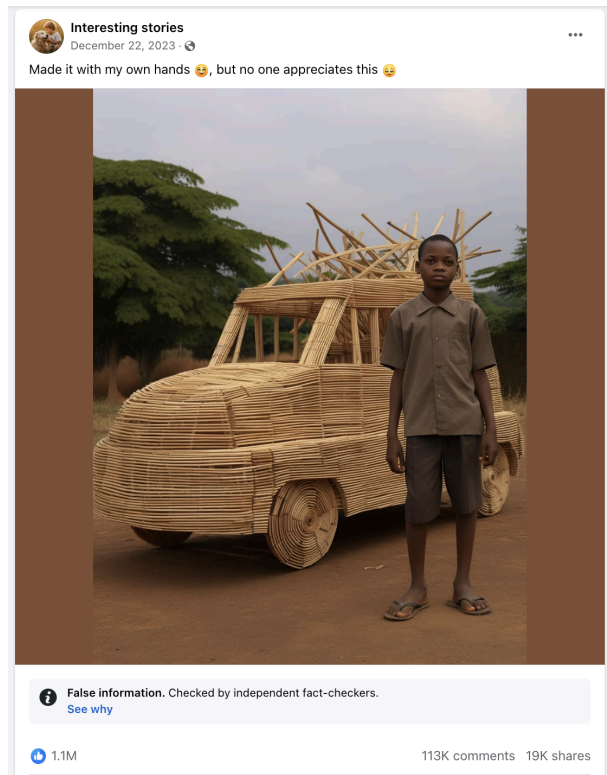

*Figure 9. One of the photos from "Interesting Stories" that received more than a million engagements was fact-checked by Congo Check and labeled by Facebook as "False information."*

## Methods

Our research required surfacing Pages, and determining whether they used large numbers of unlabeled AI-generated images. We surfaced Facebook by 1) searching for Pages using 'copypasta' captions, 2) identifying signs of coordination of those Pages with others, 3) looking at Pages that Facebook Users had called out for posting unlabeled AI-generated images, and 4) surfacing new leads from our own Facebook Feeds. We made determinations about whether Pages were using AI-generated images by finding errors in images and by analyzing trends in posts for high numbers of AI house style images. We describe our processes for surfacing Pages and making determination about use of AI below.

First, we noticed Pages posting unlabeled AI-generated images would, at times, use overlapping themes with heavy overlap or copy-and-pasted repeated captions. Searching these captions or phrases from the captions in CrowdTangle surfaced other Pages that had posted the same content in captions.

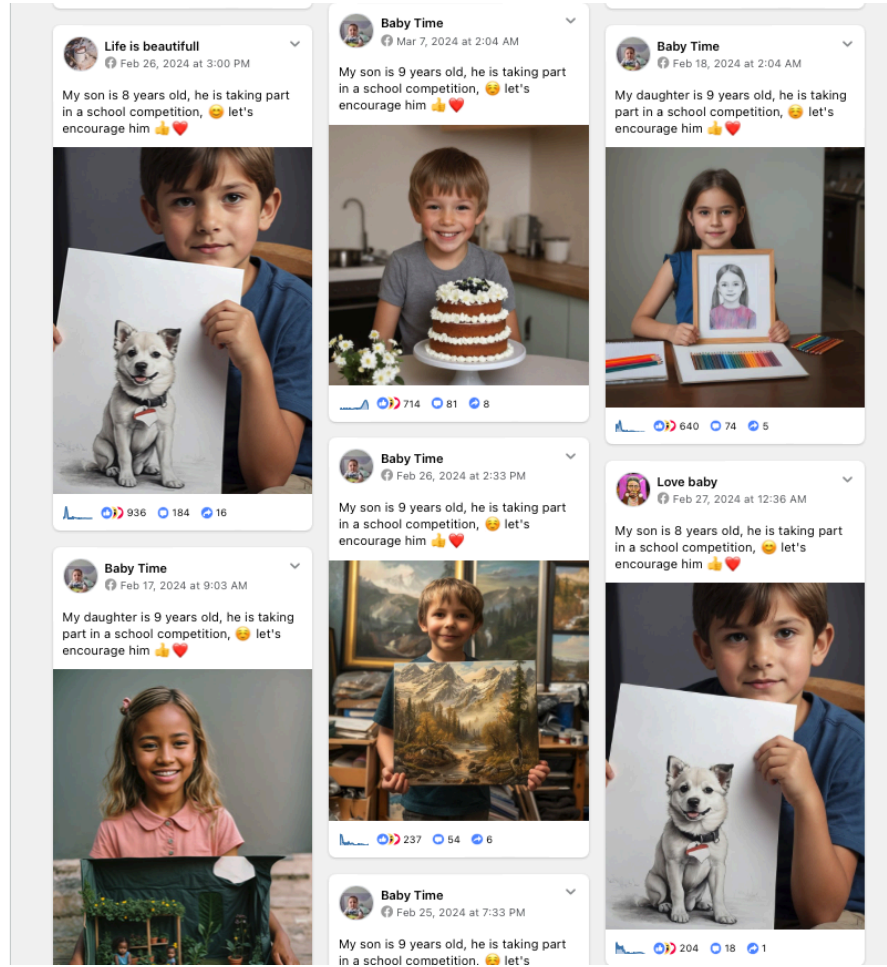

*Figure 10. Examples of Pages using overlapping phrases.* When we found a Page posting an AI-generated image, we surfaced other Pages also posting AI-generated images by searching for captions with similar phrases, such as "let's encourage him."

Second, once we had identified a Page for inclusion, we investigated adjacent Pages for signs that they also used AI-generated images en masse. For example, we looked at Pages sharing each others' content, co-owners of Groups, and Pages suggested by Facebook when viewing another. Third, we noticed Facebook groups that brought together users interested in finding AI-generated images on the platform. These groups often rely on common OSINT techniques. These groups provided several leads for our investigation. Fourth, after several days of engaging with material obtained through these searches, we began to observe unlabeled AI-generated images recommended to us on our own Feeds. In other words, searches that returned a high volume of AI-generated images across many different themes —AI-generated homes, rooms, furniture, clothing, animals, babies, people, food, artwork—returned more AI-generated images across additional random themes.

Once we surfaced a Page, we had to make a determination about whether content was AI-generated. To make this assessment, we relied on obvious mistakes or unrealistic images, as well as analyzing trends in posts. In Figure 11, we show images posted by the Page "Amazing Statues" with three hands (left), hands melded together (middle), and gloves with more than five fingers (right).

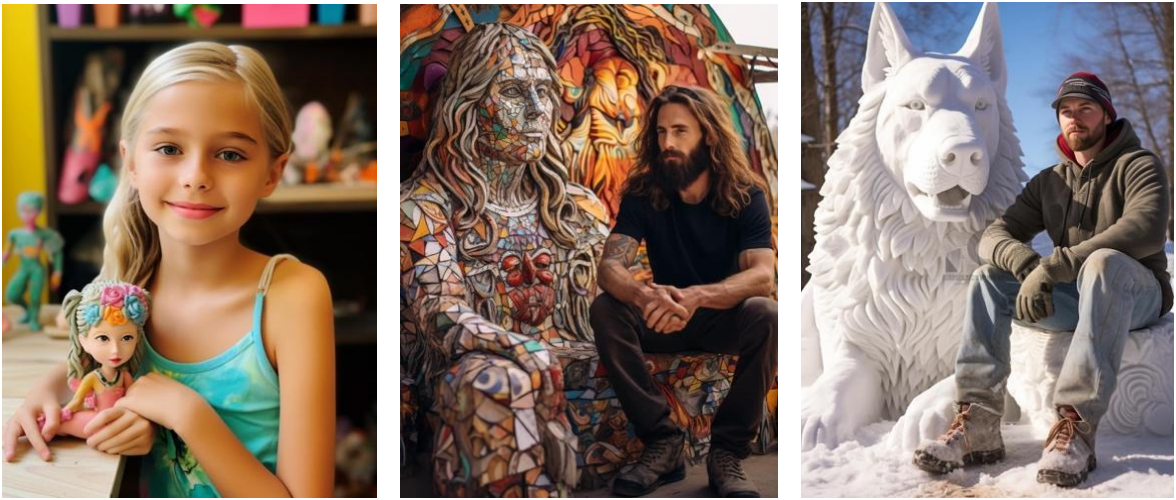

*Figure 11: Several images from the Facebook Page "Amazing Statues" with unrealistic hands. The image on the left has three hands, the image in the middle has hands that merge together, and the image on the right shows a hand with five fingers (without a thumb in view).*

We also analyzed patterns in posts: if a Page used a single AI-generated image it did not qualify, but if it posted a large number of photos (50+) that shared a house style (e.g. of Midjourney) we included it in our analysis. Just as Picasso had his Blue Period, the Pages would often go through periods, a few dozen snow carvings; a few dozen watermelon carvings; a few dozen wood carvings; a few dozen plates of artistically arranged sushi—each with a highly similar style. In Figure 12, we provide a screenshot of one such Page moving through different periods.

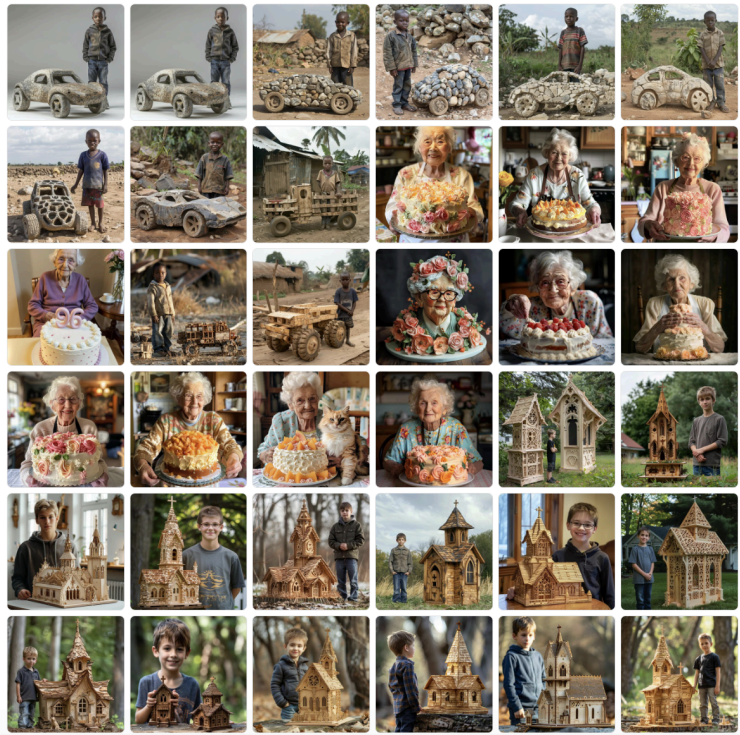

*Figure 12: Photos posted by the Page "The Amazing News." The Page posted likely AI-generated photos in batches from different genres, from children with wooden sculptures to women with cakes to children with cars made from rocks. The images have signs they are AI-generated, such as glasses that blend into wrinkles and sandals that merge into feet.*

Although these manual detection methods were sufficient for identifying Pages described above, this research method has clear limitations. It likely helped surface Pages that did not take great precautions to weed out erroneous AI-generated images or intersperse them sufficiently with real images. Even in cases where there were obvious tells, identifying those images as AI-generated still requires a high degree of cognitive effort, which may not be realistic to expect of the average Facebook user simply scrolling through their timeline. It also risks surfacing false positives, and will likely be unreliable in the future, as AI image models improve further.

This study has limitations in regards to representativeness and assessing the goals of Facebook Page operators. The study of Pages posting large numbers of unlabeled AI-generated images is neither exhaustive nor representative of all Facebook Pages. We discovered Pages that formed clusters, relied on 'copypasta' captions, or were recommended in our Feeds. This is sufficient for documenting an understudied type of misuse (and is characteristic of online investigations), but is not reflective of how unlabeled AI-generated images are used on Facebook more broadly. Second, since we do not operate the Pages ourselves nor did we interview Page operators, we cannot be sure of their aims. In some cases, their aims seemed obvious (e.g. when they posted links to the same off-platform website on many posts). At other times, the posting pattern seemed designed with the proximate goal of audience growth but an unknown ultimate goal. We encourage future research on the motivations of Page operators sharing AI-generated content, user expectations around synthetic media, and longitudinal investigations examining how those evolve.

PREPRINT DRAFT: NOT PEER REVIEWED 15Hughes, H. C., & Waismel-Manor, I. (2021). The Macedonian Fake News Industry and the 2016 US Election. *PS: Political Science & Politics, 54*(1), 19–23. https://doi.org/10.1017/S1049096520000992

Jakesch, M., French, M., Ma, X., Hanckock, J., & Naaman, M. (2019, May 2). AI-Mediated Communication: How the Perception that Profile Text was Written by AI Affects Trustworthiness. *Proceedings of the 2019 CHI Conference on Human Factors in Computing Systems.* https://dl.acm.org/doi/pdf/10.1145/3290605.3300469

Koebler, J. (2023, December). Facebook Is Being Over With Stolen, AI-Generated Images That People Think Are Real. *404 Media.* https://www.404media.co/facebook-is-being-overrun-with-stolen-ai-generated-images-that-people-think-are-real/

Koebler, J. (2024, January 8). 'Dogs Will Pass Away': Hackers Steal Dog Rescue's Facebook Page, Turn It Into AI Content Farm. *404 Media.* https://www.404media.co/dogs-will-pass-away-hackers-steal-dog-rescues-facebook-page-turn-it-into-ai-content-farm/

Koebler, J. (2024, March 19). Facebook's Algorithm Is Boosting AI Spam That Links to AI-Generated Ad Laden Click Farms. *404 Media*. https://www.404media.co/facebooks-algorithm-is-boosting-ai-spam-that-links-to-ai-generated-ad-laden-click-farms/

Ma, A., Wong, W., Bridges, C., & Concannon, K. (2023, December 12). Are the products in your shopping cart real? *The Indicator from Planet Money.* https://www.npr.org/2023/12/12/1197958902/the-indicator-from-planet-money-ai-ecommerce-12-12-2023

Maiberg, E. (2024, February 27). Ghost Kitchens Are Advertising AI-Generated Food on DoorDash and Grubhub. *404 Media*. https://www.404media.co/ghost-kitchens-are-advertising-ai-generated-food-on-doordash-and-grubhub/

Metaxas, P. T., & DeStefano, K. (2005, May). Web Spam, Propaganda and Trust. *First International Workshop on Adversarial Information Retrieval on the Web*. pp. 70-78. https://airweb.cse.lehigh.edu/2005/metaxas.pdf

Mouton, C., Lucas, C., & Guest, E. (2024). *The Operational Risks of AI in Large-Scale Biological Attacks.* RAND Corporation. https://doi.org/10.7249/RRA2977-2

Munich Security Conference. (2024, February). *A Tech Accord to Combat Deceptive Use of AI in 2024 Elections*. https://securityconference.org/en/aielectionsaccord/accord/

Phua, J. & Ahn, S. J. (2016). Explicating the 'like' on Facebook brand pages: The effect of intensity of Facebook use, number of overall 'likes', and number of friends' 'likes' on consumers' brand

PREPRINT DRAFT: NOT PEER-REVIEWED                                                                16outcomes. *Journal of Marketing Communications 22*(5), pp. 544-559. https://doi.org/10.1080/13527266.2014.941000

Roozenbeek, J., Van Der Linden, S., Goldberg, B., Rathje, S., & Lewandowsky, S. (2022, August). Psychological inoculation improves resilience against misinformation on social media. *Science Advances 8*(34). https://www.science.org/doi/pdf/10.1126/sciadv.abo6254

Seger, E. et al. (2020). *Tackling threats to informed decision-making in democratic societies: Promoting epistemic security in a technologically-advanced world.* Alan Turing Institute. https://doi.org/10.17863/CAM.64183

Spitale, G., Biller-Andorno, N., & Germani, F. (2023, June 28). AI model GPT-3 (dis) informs us better than humans. *Science Advances, 9*(26). https://www.science.org/doi/10.1126/sciadv.adh1850

Subramanian, S. (2017, February 15). The Macedonian Teens Who Mastered Fake News. *Wired.* https://www.wired.com/2017/02/veles-macedonia-fake-news/

Watukalusu, H. (2024, January 28). *Engagement-bait : Une photo générée par l'I.A faussement légendée incitant les internautes à commenter.* Congo Check. https://congocheck.net/engagement-bait-une-photo-generee-par-li-a-faussement-legendee-incitant-les-internautes-a-commenter/?fbclid=IwAR2oGPAh63Sm8_CGRa3yqOL1g81kK7qdsP1yaprIhtNtZR0avwYvDp9ZApc

Weidinger, L., Uesato, J., Rauh, M., Griffin, C., Huang, P. S., Mellor, J., ... & Gabriel, I. (2022, June). Taxonomy of risks posed by language models. *Proceedings of the 2022 ACM Conference on Fairness, Accountability, and Transparency,* pp. 214-229. https://doi.org/10.1145/3531146.3533088

PREPRINT DRAFT: NOT PEER REVIEWED 17Not applicable

**Acknowledgements**
We thank Abhiram Reddy for excellent research assistance. For feedback on our investigation or an earlier draft of this paper, we thank Elena Cryst, Shelby Grossman, Jeff Hancock, Justin Hileman, Ronald Robertson, and David Thiel.

**Authorship**
The authors made equal contributions to this research.

**Funding**
No funding has been received to conduct this research.

**Competing interests**
The authors declare no competing interests.

**Ethics**
We relied exclusively on publicly available data and did not seek IRB approval.

## Appendix: A

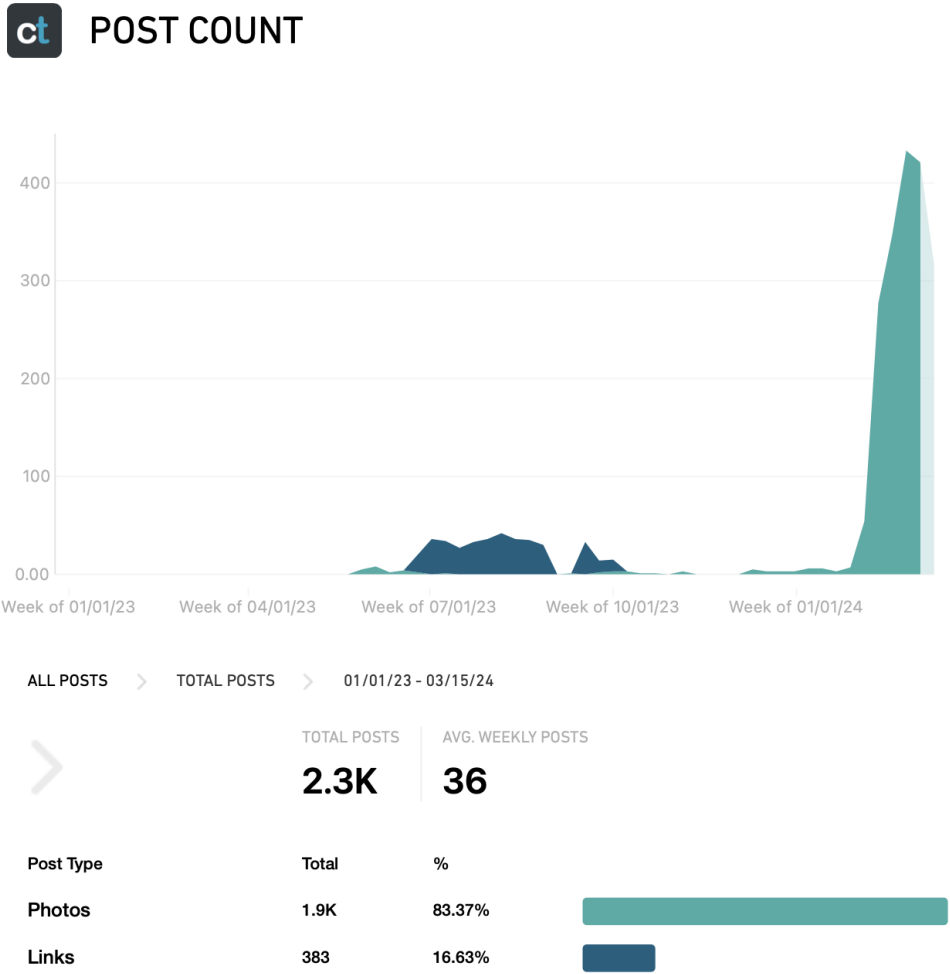

***Figure 1. Screenshots from CrowdTangle showing the number of posts that include links or photos for the Facebook Page "Love God &God Love You" over time.*** *The Page shifts from primarily posting links to posting a large number of photos.*